\documentclass{article}

\begin{document}

\begin{center}
{\Large \textbf{Jets (relativistic and non) in astrophysics}\footnote{Originally published in Italian on the popular science magazine \emph{Le Stelle}, n. 82, March 2010.}}\\
\vskip 12pt
Let's take stock of the situation on one of the most studied astrophysical phenomena during the latest years: the jets escaping from protostars, stellar singularities, GRB and active galactic nuclei.
\\
\vskip 12pt
\emph{Luigi Foschini\\
INAF - Osservatorio Astronomico di Brera, Italy}\\
\normalsize
\end{center}

When one speaks about spacetime singularities, the metaphorical term \emph{black holes} commonly used forces to think immediately to something swallowing everything and that cannot allow anything to escape. Instead, the nature offers us very different shows, where enormous structures of plasma (jets) expands at speeds close to that of light (said \emph{relativistic}) from the singularities toward the outer space, forming structures with lengths up to about one megaparsec (more than 3 millions light years). These jets are very common among the cosmic sources: from the relativistic jets in the stellar or supermassive singularities, in the gamma-ray bursts (GRB) and in the neutron stars, up to the supersonic jets in the protostars. One can expect that any object accreting material onto its equatorial plane, could develop also polar jets. However, how and why this can happen is still almost unknown.

The observational evidences are now a lot and date back to the dawn of the radio astronomy in the fifties, even though the first jet to be discovered was that of the radio galaxy M87 observed by an optical telescope. Indeed, already in 1918, the American astronomer Heber Curtis (mostly known because of the \emph{Great Debate} with Harlow Shapley on the nature of nebulae), while observing at the Lick Observatory, described a curious protuberance extending from the nucleus of M87. During fifties, with the advent of the first radio interferometers, it was possible to outline the first extended structures of the most intense radio sources. However, the early interpretations of these so unusual objects were focused on the possibility to observe colliding galaxies (as it occurred in the case of the radio galaxy Cygnus A).

In 1966, a twenty-four-years-old Martin Rees(\footnote{M. Rees, 1966, ``Appearance of Relativistically Expanding Radio Sources'', \emph{Nature}, 211, 468.}) suggested that the rapid intensity variations of the radio emission of these extragalactic sources could be due to a change of dimensions (expansion) occurring at relativistic speed. He applied his theory to 3C 273, a quasar discovered a few years before, and the calculations matched the observations. In the following years, with the improvements and expansion of the radio observatories, the number and the type of these objects increased. Then, X- and $\gamma-$ray observations from satellites were added. Particularly, during nineties, the launch of the US satellite \emph{Compton Gamma-Ray Observatory} allowed the discovery of high-energy $\gamma$ rays ($E>100$ MeV) from these sources. 

However, there was a real \emph{zoo} of sources, mainly due to the fact that the initial classification was assigned on the basis of the instrument used for the discovery. For example, certain quasars optically discovered were classified as \emph{Optically Violent Variables} (OVV), while if they were discovered by radio telescopes, they were indicated as flat-spectrum radio quasars (FSRQ). But it simply was the same type of source observed in one frequency band or in another. 

In 1995, when the amount of observations was sufficiently mature, Megan Urry and Paolo Padovani(\footnote{C.M. Urry \& P. Padovani, 1995, ``Unified Schemes for Radio-Loud Active Galactic Nuclei'', \emph{Publications of the Astronomical Society of the Pacific}, 107, 803.}) reviewed a unified model for this type of sources (the basis were set up already in 1978 at the Pittsburgh Conference -- see later in the text), which is still roughly valid today. According to this model, these sources are \emph{active galactic nuclei} (AGN) made of a supermassive spacetime singularity of the order of billions of solar masses, accreting matter from the host galaxy and ejecting a bipolar jet at relativistic speed. The presence of the singularity casts the nearby spacetime and, going from the center to the outskirts, it is possible to see:

\begin{enumerate}

\item \emph{accretion disk}, made of the matter attracted by the singularity and forming a kind of doughnut on the equatorial plane;

\item \emph{broad-line region} (BLR), a region of plasma emitting photons at specific frequencies (\emph{emission lines}), which in turn depends on the chemical elements composing the gas; to hold the orbit at distances from the singularity typical of BLR ($\approx 0.1-1$ pc), it is necessary that the plasma is moving at speeds of several thousands of kilometers per second and this determines a broadening of the profile of these emission lines, from which the name of the region;

\item \emph{molecular torus}, another kind of equatorial doughnut made of cold matter, which can reach densities so high to absorb X-ray photons of energies up to a few keV;

\item \emph{narrow-line region} (NLR), a region of plasma similar to the BLR, but being at larger distances from the central singularity ($\approx 10-100$ pc), it requires smaller orbital speeds (a few hundreds of kilometers per second) and, therefore, the line profiles are narrower, from which the name of the region.

\end{enumerate}

It is possible to observe different types of sources depending on the viewing angle: if the angle is very large, up to a view along the equatorial plane, then the source is called \emph{radio galaxy}. If the viewing angle is very small, just a very few degrees, then the source is named \emph{blazar}, contraction of \emph{BL Lac} and \emph{quasar}. This name was proposed by Ed Spiegel in 1978, during the well-known conference on BL Lac objects held in Pittsburgh, on the basis of the seminal talk by Roger Blandford and Martin Rees(\footnote{R.D. Blandford \& M.J. Rees, 1978, ``Some comments on radiation mechanisms in Lacertids'', In: \emph{Proceedings of the Pittsburgh Conference on BL Lac Objects}, Pittsburgh, 24-26 April 1978, University of Pittsburgh, p. 341.}). In their talk, these two scientists underlined the similarities of the electromagnetic emission of some quasars and BL Lacs, while in the discussion after the talk, emerged the possibility of a link with the radio galaxies by taking into account the viewing angle. Practically, the basis of the unified model, later systematized by Urry \& Padovani in 1995, were set up. 

In 1998, a group of Italian researchers (Celotti, Comastri, Fossati, Ghisellini, Maraschi, in alphabetical order) analyzed a sample of blazar and concluded that these sources form a sequence, today known as the \emph{blazar sequence} (\footnote{G. Fossati et al., 1998, ``A unifying view of the spectral energy distributions of blazars'', \emph{Monthly Notices of the Royal Astronomical Society}, 299, 433; G. Ghisellini et al., 1998, ``A theoretical unifying scheme for gamma-ray bright blazars'', \emph{Monthly Notices of the Royal Astronomical Society}, 301, 451.}). In this sequence, the frequencies, where is recorded the maximum emission of the electromagnetic radiation, depends on the power emitted by the jet. The broad-band electromagnetic spectrum (i.e. the emitted flux as a function of the frequency, from radio to $\gamma$ rays) of the AGN with relativistic jets is characterized by two humps, one at low frequency -- in the region including infrared to soft X-rays -- while the other is in the band of $\gamma$ rays. The first hump shows the emission characteristics of the synchrotron radiation, i.e. the emission from relativistic electrons moving along the magnetic field lines. The rotation is an accelerated motion, characterized by a continuous change of direction and, as known from basic physics, an accelerated electric charge emits electromagnetic radiation. Instead, the second hump is explained with different theories, although the most reliable one illustrates this emission as due to the same population of relativistic electrons that hit low-energy photons (\emph{seed photons}), transferring a part of their energy. This process is called \emph{inverse-Compton effect}, after Arthur H. Compton who discovered it in 1922 and received the Nobel prize for physics in 1927. The seed photons can be the same of the synchrotron (in this case, it is named \emph{synchrotron self-Compton}, SSC) or can come either from the accretion disk or the BLR or the molecular torus (in this case, it is named \emph{External Compton}, EC). By absorbing part of the electron energy, the photons can reach the energies typical of $\gamma$ rays (MeV-TeV). 

In the blazar sequence, the quasars, which have the most powerful jets, have the emission peaks at relatively low frequencies (infrared for the synchrotron and MeV-GeV $\gamma$ rays for the inverse-Compton). On the other hand, the BL Lac sources, with lower power, have the peaks at relatively high frequencies (ultraviolet/X-rays for the synchrotron and TeV $\gamma$ rays for the inverse-Compton). 

The radiation emitted by the jet in blazars overwhelms any other emission from nearby structures of the AGN. This occurs because of the effects of special relativity: indeed, when a source of photons travels at speed close to that of light, the photons arrange themselves along a cone centered on the source and whose aperture depends on the speed. The cone is narrower as the speed of the source is closer to that of light. This phenomenon, somehow similar to the rain drops falling obliquely on a runner, is called \emph{relativistic beaming}. The effects on the emission of the jet are to amplify the luminosity and to shorten the variability time. Therefore, the jet emission, particularly when it is very active, is stronger than that of the other components. When the jet is quiet or viewed at larger angles, it is possible to observe the emission from other structures, such as the accretion disk. 

The relativistic beaming has another important effect: it reduces the probability of interaction between photons, which in turn is the key for the emission of high-energy $\gamma$ rays. Indeed, $\gamma$ rays with energies greater than about $1$ MeV are equivalent to one electron-positron pair ($e^{+}e^{-}$), according to the famous Einstein equation $E=mc^2$, because the rest mass of each particle is $511$~keV ($2\times 511$~keV~$=1022$~keV~$=1.022$~MeV). However, one photon with such an energy does not decade spontaneously into an $e^{+}e^{-}$ pair, but it needs of a third particle to obey to the momentum conservation law. The probability of interaction of a photon with other particles is maximum in the case of another photon and, therefore, the high-energy photons of a jet could be transformed in $e^{+}e^{-}$ pairs by interacting with the photons of the nearby structures (accretion disk, BLR, molecular torus, NLR, host galaxy stars, ...). Since it is not possible to observe electrons and positrons from very far sources, the information would be lost. The relativistic beaming, by reducing the probability of photon-photon interaction, allows to high-energy photons to escape from the AGN and reach us. 

This phenomenon is very important in the studies of astrophysical jets: indeed, although jets were known since seventies, because of radio observations, it was possible to elaborate a reliable jet model only in nineties thanks to the \emph{Compton Gamma-Ray Observatory}, which discovered that the high-energy $\gamma-$ray sky is dominated by the blazars emitting most of their power in this energy band. 

The present paradigm of the relativistic jet can be summarized as follows: there is a magnetic field which supplies the rails on which electrons are put and accelerated at relativistic speed. These electrons emit synchrotron radiation while moving along the magnetic field lines and transfer part of their energy when they hit with low energy photons, either their synchrotron photons or photons coming from nearby structures. If one observes the jet with a small viewing angle with respect to the direction of motion (blazar), then the luminosity is amplified and the variations of emission are extremely rapid. Instead, if the viewing angle is large (radio galaxies), then the relativistic effects are smoothed and it is possible to observe the emission from other structures. Particularly, it is though that the jet is composed of a central fast spine and a slower external layer(\footnote{G. Ghisellini et al., 2005, ``Structured jets in TeV BL Lac objects and radiogalaxies. Implications for the observed properties'', \emph{Astronomy and Astrophysics}, 432, 401.}). As the viewing angle increases, the luminosity of the central spine of the jet becomes weaker and weaker, but in turn the $\gamma-$ray emission due to the spine-layer interaction starts increasing. This theory has not yet been tested, but the \emph{Fermi} satellite now in orbit will give the necessary data for this purpose. Anyway, it is ascertained that the viewing angle has a role of paramount importance in the AGNs with relativistic jets.

This scheme included also another thing: the host galaxy, which always was of elliptical type both for blazars and radio galaxies, while AGNs without jets were hosted by spirals. Therefore, it was thought about the possibility of a link between the host galaxy and the presence of a jet. However, in 2009, the satellite \emph{Fermi} measured the emission of high-energy $\gamma$ rays typical of relativistic jets from a class of sources named Narrow-Line Seyfert 1 (NLS1), a kind of AGN generally without jets and hosted in spiral galaxies(\footnote{A.A. Abdo et al., 2009, ``Radio-Loud Narrow-Line Seyfert 1 as a New Class of Gamma-Ray Active Galactic Nuclei'', \emph{The Astrophysical Journal}, 707, L142.}). The jet is quite similar to that in blazars, i.e. with very small viewing angles, but the NLS1s have a BLR different from that in BL Lacs and quasars. The different BLR seems to be not influential on the jet, while the fact of a spiral host galaxy breaks down the paradigm of a link between the jet and the host galaxy. 

On the other hand, the non-existence of this link could be foreseen, as it is known that jets develop in the most disparate conditions. Indeed, already at the beginning of nineties, it was known the existence of relativistic jets also in other types of sources, like the Galactic binary systems. In 1992, I.F. Mirabel, L.F. Rodriguez, B. Cordier, J. Paul and F. Lebrun observed with a radio telescope (VLA) a bipolar jet emitted by a compact and variable nucleus of one source close to the Galactic center, named the \emph{Great Annihilator} or 1E~$1740.7-2942$ (\footnote{I.F. Mirabel et al., 1992, ``A double-sided radio jet from the compact Galactic Centre annihilator 1E140.7 - 2942'', \emph{Nature}, 358, 215.}). It was a stellar mass black hole accreting from a companion star, which mimed at Galactic scales the phenomena observed in the AGNs with relativistic jets discovered years before. Because of these characteristics, they called it \emph{microquasar}. During the following years, many other microquasars were discovered in the Milky Way (GRS~$1915+105$ and XTE~J$1550-564$ are among the best known) and now are an accepted reality. The structure of the microquasar is a little different from that of AGNs: there is still a singularity, but with masses of the order of the solar mass (up to a few tens), accreting matter by the usual disk from a companion star, which in turn is generally with low mass. It is rare to find a blue supergiant, but not impossible (e.g. Cygnus X-1). There are no BLR, NLR or obscuring tori and the structure is very essential. Nevertheless, the relativistic jet is still present and fully developed, to become $\gamma-$ray emitter, as recently observed both by Cerenkov telescopes and by the \emph{Fermi} satellite. In this case, the seed photons for the high-energy emission are coming either from the companion star or from the synchrotron process (SSC).

Moreover, recent studies have shown that the presence of a spacetime singularity is not even necessary. Indeed, relativistic jets have been observed even from binary systems with neutron stars or white dwarfs, although less powerful than those powered by black holes(\footnote{See a review R. Fender, 2006, ``Jets from X-ray binaries''. In: \emph{Compact stellar X-ray sources}. Edited by W. Lewin \& M. van der Klis. Cambridge University Press, p. 381.}). It is worth noting that not all the neutron stars show jets, but only those with the accretion disk. Pulsars with high magnetic field, which destroys the disk, are thus excluded, although some jet-like structures have been observed in these sources. Nevertheless, this fact is of paramount importance, because it states that the jet does not depends on the presence of a singularity, although its power is strongly affected.

To confirm this, there is the observation of bipolar jets -- although not relativistic -- in protostellar systems(\footnote{See a review J. Bally, 2009, ``Jets from young stars''. In: \emph{Protostellar Jets in Context}, Edited by K. Tsinganos, T. Ray, M. Stute, Springer, p. 11.}). These systems were discovered at the end of the XIX century by S.W. Burnham and had a significant advancement during fifties thanks to the work by G. Herbig and G. Haro, but the discovery of supersonic jets is relatively recent. The accretion is given by the gas of the nebula where the protostar is forming and the jet moves itself at supersonic speed of hundreds-thousands of kilometers per second. Shock waves emitting radiation are formed when the ejected gas interacts with the interstellar matter. 

Finally, the Gamma-Ray Bursts (GRB): they were observed for the first time during sixties by the US military satellites of the series Vela, launched to monitor the atomic experiments of the URSS. Only in 1973, the data were published on an international scientific journal, thus becoming accessible to scientists. During nineties, the experiment BATSE (\emph{Burst And Transient Spectrometer Experiment}) onboard the \emph{Compton Gamma-Ray Observatory}, observed about 3000 GRBs uniformly distributed all over the sky, thus indicating an extragalactic origin of these phenomena. In 1997, the Italian-Dutch satellite \emph{BeppoSAX} succeeded -- for the first time -- to observe the \emph{afterglow}, i.e. the echo of the $\gamma-$ray explosion at lower energies (X-rays). This made it possible to find the optical counterpart and then to measure the redshift. The farthest GRB observed to date measures $z\approx 8.1-8.2$ (GRB090423) corresponding to an age of the Universe of only 631 millions of years(\footnote{N.R. Tanvir et al., 2009, ``A $\gamma-$ray burst at a redshift of $z\approx 8.2$'', \emph{Nature}, 461, 1254; R. Salvaterra et al., 2009, ``GRB 090423 at a redshift of $z\approx 8.1$'', \emph{Nature}, 461, 1258.}). 

A GRB is grossly an expanding fireball and, when it encounters the interstellar medium, it interacts with it generating high-energy $\gamma$ rays by Compton effect, where the seed photons are those of the interstellar medium(\footnote{See reviews in: G. Ghisellini \& G. Ghirlanda, 2010, ``Fermi/LAT Gamma Ray Burst emission models and jet properties'', \emph{Proceedings of the Conference The Extreme sky: Sampling the Universe above 10 keV}, Otranto, October 2009, \texttt{arXiv:1002.3377}; G. Ghisellini, 2010, ``What is the radiative process of the prompt phase of Gamma-Ray Bursts?'', \emph{Proceedings of the Conference X-ray Astronomy 2009}, Bologna, September 2009, \texttt{arXiv:1002.1784}.}). It is thought that the origin of the fireball could be due either to phenomena linked to supernovae (for long GRBs, with duration greater than a few seconds) or to the collision and merging of two neutron stars or one neutron star and one black hole (for short GRBs, with duration less than a few seconds). Although, it was never directly observed, as in AGNs or Galactic binaries, the presence of an ultrarelativistic jet in the GRBs is needed to explain with reasonable hypotheses the energy measured. It is not clear what could be the accretion mechanism: whether the usual disk around neutron stars or stellar mass singularities or the re-falling of material ejected by the supernova explosion. This lack of well-grounded indications is not a surprise, given the uncertainties on the central engine of the GRBs.

\begin{table}[ht!]
\scriptsize
\caption{\scriptsize{Summary of jet systems. Acronyms: RG = radio galaxy; NLS1 = Narrow-Line Seyfert 1; SS = spacetime singularity (black hole); NS = neutron star; WD = white dwarf; PS = Protostar; SN = supernova.}}
\begin{center}
\begin{tabular}{lcccc}
\hline
\textbf{Source} & AGN & Galactic Binaries & Protostars & GRBs \\
\hline
\textbf{Type of Jet}     & Relativistic & Mildly Relativistic & Supersonic & Ultrarelativistic \\
\textbf{System}  & blazar & Stellar SS & PS & SN\\
{}  & RG & NS & {} & NS+NS\\
{}  & NLS1 & WD & {} & SS+NS\\
\textbf{Mass}            & $\approx 10^{6-10}M_{\odot}$ & $\approx 1-20M_{\odot}$ & $<1M_{\odot}$ & a few $M_{\odot}$\\
\textbf{Accretion}       & disk & disk & disk & disk? other?\\
\textbf{Host}            & Elliptical or Spiral & Galactic Binary & PS Nebula & ? \\
{}                       & Galaxy & with star & {} & {} \\
\hline
\end{tabular}
\end{center}
\label{tab:riassunto}                                                   
\normalsize
\end{table}%
 
To conclude, in the Table~\ref{tab:riassunto}, are summarized the different types of jets, relativistic and non, today known in astrophysics. The most powerful sources are GRBs, which release enormous amounts of energy ($10^{51-54}$~erg or $10^{44-47}$~Joule) in a few seconds or even less, while blazars have slightly less power ($10^{47-49}$~erg/s or $10^{40-42}$~Joule/s), but the emission can be weeks or even months long. The present day record for GRB is held by the event 080319B ($z=0.937$, equal to $7.5$ billions of light years), which was visible by naked eye with a visual magnitude of $5.5$: it released an energy of more than $10^{54}$~erg. The record for blazars is held by 3C 454.3 ($z=0.854$ or $7.1$ billions of light years), which during the early days of december 2009 reached a power of more than $10^{49}$~erg/s; however, since it held this power for about 10 days, the released energy was more than $10^{55}$~erg. It was the most energetic event known to date after the Big Bang. 

From Table~\ref{tab:riassunto}, it can be seen that jets are quite common and independent on the system. If we exclude the GRBs, which are impulsive phenomena, it is clear than as the mass of the central object decreases, the speed of the jet also decreases. It is relativistic in AGNs, mildly relativistic in binaries and supersonic in protostellar systems. Also the accretion rate seems to affect the characteristics of the jets and another parameter of a certain importance is the self-rotation of the source (spin). How these parameters (mass, accretion rate, spin) determine the characteristics of the jets, which in turn is certainly linked on how these structures were born and develop, is still one of the greatest and most fascinating mysteries of the contemporary astrophysics.

\vskip 12pt
\emph{I would like to thank Gabriele Ghisellini for the interesting and useful comments to the text.}

\end{document}